\documentclass{vgtc}                          

\ifpdf
  \pdfoutput=1\relax                   
  \usepackage{graphicx}                
  \DeclareGraphicsExtensions{.pdf,.png,.jpg,.jpeg} 
\else
  \usepackage{graphicx}                
  \DeclareGraphicsExtensions{.eps}     
\fi%

\graphicspath{{figures/}{pictures/}{images/}{./}} 

\usepackage{microtype}                 
\PassOptionsToPackage{warn}{textcomp}  
\usepackage{textcomp}                  
\usepackage{mathptmx}                  
\usepackage{times}                     
\usepackage{cite}                      
\usepackage{tabu}                      
\usepackage{booktabs}                  

\usepackage{flushend}
  
\onlineid{0}

\vgtccategory{Research}

\vgtcinsertpkg

\title{Developing Effective Community Network Analysis Tools
According to Visualization Psychology}

\author{Darren J. Edwards\thanks{e-mail: d.j.edwards@swansea.ac.uk}\\ %
        \scriptsize Swansea University, UK %
\and Min Chen\thanks{e-mail: min.chen@oerc.ox.ac.uk}\\ %
     \scriptsize University of Oxford, UK}

\teaser{
  \includegraphics[width=160mm]{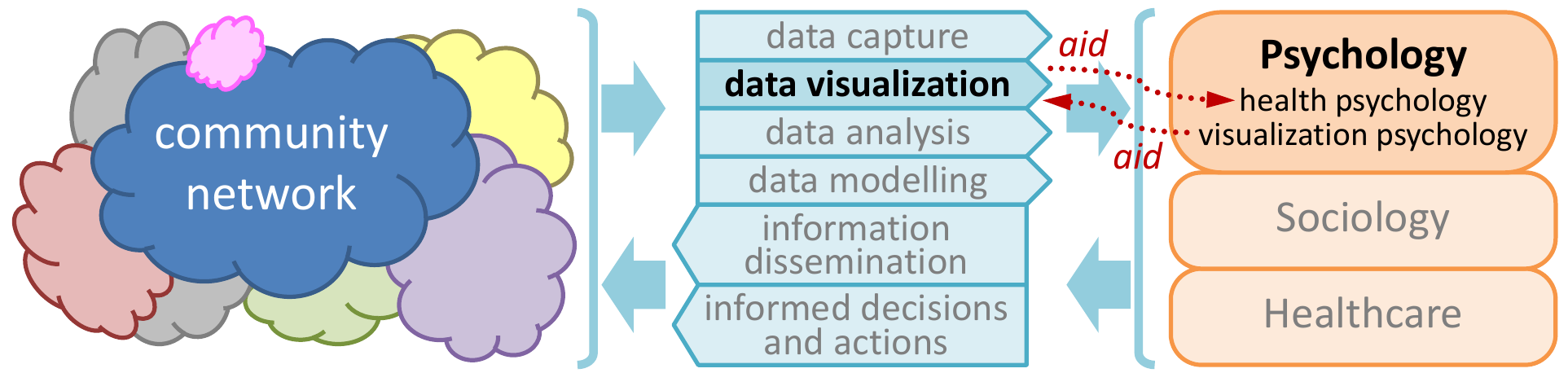}
  \caption{Data visualization can aid community network analysis in health psychology, while visualization psychology can aid the improvement and optimization of data visualization processes.}
  \label{fig:Draw}
}

\begin{document}


\firstsection{Introduction}

\maketitle


Community network analysis is advancing in areas of sociology and geography, but remains largely undiscovered in the realm of health psychology. The outbreak of COVID-19 around the world highlights the critical roles that community networks can play in social and psychological health and wellbeing. Network visualization is an energetic subarea of visualization. It is an indispensable technology for supporting community network analysis, aiding psychologists of all backgrounds in examining community organizational effects on wellbeing. Meanwhile, visualization psychology is a new branch of applied psychology\footnote{To our best knowledge, the term of ``visualization psychology'' was first suggested in \cite{Abdul-Rahman:2019:arXiv,Abdul-Rahman:2019:survey} and the \emph{IEEE VIS 2020 Workshop on Visualization Psychology} signified the establishment of this new subject.}, which aims to study the effects of visualization on different stakeholders in data intelligence workflows. As illustrated in Figure \ref{fig:Draw}, visualization psychology can, in return, aid visualization scientists to design better visualization tools for the practitioners in all disciplines in general and those in health psychology in particular. In this paper, we reflect on the importance and potential impact of developing visualization psychology in the context of community network analysis.


The recent developments of social networks have shaped a new global ecosystem where individuals are linked more closely, and their impact on wellbeing has been recognized \cite{Christakis:2009:book}. In social neuroscience, it has been observed that specific areas of the brain, which are activated on the basis of social connection, link the need for connection with wellbeing on one level \cite{Lieberman:2013:book}. On another level, this contrasts with diminished values, connection to others, particularly at a societal level, which may have been brought about by the acceptance of neo-liberal ideals, promoting individualism, and self-promotion \cite{Gruebner:2017:DAI}. There is growing evidence of the individuals, and their relationship with their environment, to explain causal functions of wellbeing \cite{Mortal:2015:L,Mead:2019:book,Kemp:2017:book}.  Given this, then a greater methodological drive for evidencing such interrelated connections are perhaps needed, and within the areas of health, social, clinical, and environmental psychology.

\section{Community Network Analysis: A Discussion from the Perspective of Visualization Psychology}
In community network analysis, commonly based on graph theory, analytical algorithms are used to identify possible connections among individuals and different community groups \cite{Radicchi:2004:PNAS}. There are numerous data mining metrics for determining if two individuals are related and should be connected in a community network. One of such metrics utilizes Cohen's $\kappa$ coefficient, which is a similarity measure for categorical data and has been traditionally applied to data mining in the form of cluster analysis, network link prediction, and in understanding how the mind manages information in categorization research \cite{Hoffman:2018:MBR}. The resultant networks are typically depicted using node-link diagrams for disseminating the analytical results to the public or informing the relevant decision makers.

In healthcare and health psychology, such a data intelligence workflow can aid community healthcare professionals in (i) monitoring and managing mental wellbeing in the society, (ii) conducting various operations in epidemic management (e.g., contact tracing, risk assessment, and intervention planning), and evaluating and optimizing various theory-based and data-driven models. While visualizing and exploring such analytically-constructed community networks can be very useful, there are many unanswered questions in visualization psychology, which suggest that the above workflow from data to analysis, then visualization, and finally decision may not be as dependable as it appears. For example, these questions may concern the following cognitive aspects:

\begin{enumerate}
    \vspace{-1mm}
    \item[a.] \textbf{Trust.} When a potential connection between two individuals is depicted as a solid line, it may induce a sense of trust of the analytical metric used to determine the connection. In many practical situations, such a metric can be rather unreliable. Many applications in visual analytics (a sub-field of visualization) have shown that the chance of false positive and false negative can be rather high in some situations.
    \vspace{-1mm}
    \item[b.] \textbf{Spatial Reasoning.} Because human heuristics instinctively use spatial distance to reason about closeness, the proximity between nodes in a node-link diagram often misleads some viewers to perceive the level of similarity or association between nodes. It is not known whether it demands some extra cognitive effort to suppress such instinct.
    \vspace{-1mm}
    \item[c] \textbf{Working Memory.} The changes of a network are commonly visualized with animation, which is known to be inefficient for supporting in-depth observation and analysis because of the limited working memory capacity.
    \vspace{-1mm}
    \item[d.] \textbf{Visual Search.} Forced directed graph drawing is a popular visualization technique, and many viewers are impressed by the animated convergence from an arbitrary layout to a layout based on the edge weights. However, because the spatial locations of nodes are unstable, it usually demands extra cognitive effort in cases where users need to perform visual search tasks.%
    \vspace{-1mm}
    \item[e.] \textbf{Motivation.} When modellers using visualization for depicting simulation results, there is often a “mental conflict” between using visualization to convince others to accept the prediction by the model, and using visualization to evaluate the model and validate the findings.
    \vspace{-1mm}
    \item[f.] \textbf{Confidence.} Many professionals in healthcare often felt inadequate in front of analytical conclusions resulting from mathematical modelling, and may sometimes be prevented from using their knowledge and initiatives in operation management. One may hypothesize the benefit of using visualization in evaluating and scrutinizing such analytical conclusions (see also (e)), or on the contrary, the shortcoming of using visualization if it may induce unjustifiable trust (see also (a)).
\end{enumerate}

There are numerous phenomena in the field of visualization and many visualization applications are waiting to be studied in visualization psychology. Meanwhile, with effective visualization, we can address many challenges in health psychology, including:

\begin{enumerate}
    \item In the context of epidemic management, (i) advising the current state of social engagement of individual persons for whom medical and healthcare professionals have recommended to increase and maintain a certain level/form of social engagement for their well-being; (ii) improving cognitive effectiveness and efficiency in monitoring process, and addressing issues such as inattentional blindness, limited working memory, poor collaboration and communication between healthcare workers.
    \item In community care management, to enable individuals to observe their own social engagement, in order to, e.g., motivate them to increase or decrease a certain type of social engagement. The following may be important: (a) being motivated by others who are in a similar situation; (b) being able to reach out to others who are in a similar situation or can offer help; (c) feeling rewarded by adhering to a treatment plan through a community network. The cognitive considerations include emotion and motivation as well as many aspects of social psychology.
\end{enumerate}

\section{Conclusions}
In conclusion, it is highly desirable for medical and healthcare professionals to be able to use visualization as part of their data intelligence workflows. The benefits of visualization are evident in many other applications. On the other hand, the uses of visualization techniques in healthcare and health psychology can benefit from research in visualization psychology, which will facilitate the development of more effective and efficient visualization techniques. Information theory has provided explanation why visualization can complement the limitation of machine-centric processes (e.g., statistics, algorithms, and machine learning) in data intelligence workflows \cite{Chen:2016:TVCG}. Cognitive studies have shown that visualization empower human viewers to use their knowledge to make up for the shortcomings of machine-centric processes, e.g., \cite{Kijmongkolchai:2017:CGF}. It has been postulated that the mathematical rationale for visualization may be applicable to many phenomena in psychology \cite{Chen:2020:book}.

However, there are not yet many empirical studies on advanced visualization methods in the context of healthcare. There has already been many visualization applications in healthcare (e.g., \cite{Preim:2020:CGF,Zakkar:2017:OJPHI}).
There is an urgent need for such research, which can help deploy visualization techniques in healthcare through rigorous scientific processes, while improving out scientific understanding in visualization psychology.
While we should continue to develop new visualization tools for healthcare professionals, including health psychologists, in parallel, we should make a serious effort to understand the cognitive benefits and shortcomings of using forms of visualization.
In many ways, health psychologists are in a unique position, where they have the mandate to deploy visualization techniques, the opportunities to collect observational data about the effectiveness of visualization, and the knowledge to relate their observation to the fundamental theories in psychology.

In conclusion, because visualization applications in healthcare are typically 
risk-averse, health psychologists can play a significant role in ensuring appropriate and effective uses of visualization techniques in healthcare.
Health psychologists are thus at the center of the triangle of ``health science'', ``visualization technology'', and ``visualization psychology'', and can enable the two-way knowledge flow depicted in Figure \ref{fig:Draw}.



\bibliographystyle{abbrv-doi}
\bibliography{references}
\newpage
\end{document}